# The Accuracy of IRI in the Description of Development of the Equatorial (Appleton) Anomaly

Yurii V. Dumin

*Institute of Terrestrial Magnetism, Ionosphere, and Radiowave Propagation, Troitsk, Moscow reg. 142190, Russia.  E-mail: dumin@yahoo.com*

**Abstract**

   An accuracy of the last version of International Reference Ionosphere IRI-2001 at low latitudes is tested by a detailed comparison of longitudinal variations in the equatorial anomaly development following from this model with the ones given by independent ISS-b satellite data.  The critical frequency of F2-layer was calculated by either CCIR or URSI submodel, both of which are incorporated into IRI-2001.  The main conclusion of our analysis is that the longitudinal variations given by CCIR are substantially oversmoothed in comparison with the ISS-b satellite data, and the variations given by URSI are further smoothed with respect to the CCIR ones.  Therefore, further improvements of model description of the low-latitude F2-layer are desirable.

*Key words:*
Low-latitude F2-layer, Equatorial (Appleton) anomaly, Longitudinal variations, International Reference Ionosphere, IRI-2001, URSI, CCIR, ISS-b

## 1. Introduction

   As is known, a worldwide distribution of the electron concentration in the F2-layer maximum (commonly characterized by the critical frequency $f_0$F2) is one of cornerstones of IRI, because it serves as basis for constructing height profiles of all other ionospheric parameters (Rawer, 1984a, 1984b; Rawer et al., 1978).  So, the accuracy of presentation of $f_0$F2 is an important criterion of overall consistency of this model.
   The work on worldwide mapping of $f_0$F2 began long before the establishment of IRI (e.g. review by Bradley, 1990).  The main problem here is a strongly-nonuniform distribution of the available data of ionospheric sounding over the Earth's regions (namely, their insufficient amount in the southern hemisphere and absence in the extensive oceanic areas), which leads to serious obstacles in the numerical interpolation procedure.  They are usually got around by introduction of a set of the so-called "screen" points, in which $f_0$F2 are taken as average values derived from the data available for the same latitudes.[1]  As a result, the interpolation procedure becomes well-posed, but longitudinal behavior of the critical frequency turns out to be inevitably distorted.  This problem was discussed in many previous IRI meetings (Bilitza, 1995; Bilitza et al., 1987, 1993; Ramanamurty, 1984; Rawer, 1984c, 1994), and much efforts have been undertaken in the last two decades to improve the maps of $f_0$F2 incorporated into IRI.
   The aim of the present report is to review a current situation for the particular case of the low-latitude ionospheric F2-layer.  As is known, its most important morphological feature is the equatorial anomaly.  In the original work by Appleton (1946) this anomaly was identified as the critical frequency distribution with minimum at the magnetic equator and two maxima on both sides from it (at the magnetic latitudes about ±15°).  Later, starting approximately from the International Geophysical Year (1957–1958), it was found that in some cases there was only one well-developed crest, substantially shifted from the magnetic equator; and this more general phenomenon was usually called the equatorial (rather than Appleton) anomaly.  Nevertheless, our subsequent consideration will be concentrated on the "classical" Appleton anomaly, where both crests are present.

---

[1] In fact, the effective latitude, corrected for the magnetic inclination, is commonly used in such calculations.  More details for IRI can be found in Rawer (1984a); and for a similar model developed in Russia, in Chernyshev and Vasil'eva (1973).

In testing the ionospheric models, various morphological features can be used. A reliable model has to be in agreement with the observational data by all such characteristics. A disagreement in any of them (for example, the degree of development of the two-crest Appleton anomaly) implies that the model requires improvement.

## 2. Indices characterizing the development of Appleton anomaly

First of all, we should answer the question what is the most appropriate parameter to describe the degree of development of the Appleton anomaly? In principle, any effective index introduced for this purpose is quite arbitrary. So, strictly speaking, none of them can be called "correct" or "incorrect". Nevertheless, it is reasonable to impose some natural physical requirements on its behavior. Since we are going to study the case of the two-crest equatorial anomaly, the following main requirements can be formulated:

(1) the corresponding index should increase continuously from zero when the anomaly begins to develop and decrease continuously to zero when the anomaly disappears; and

(2) this index should characterize simultaneously the amplitudes of both crests of the anomaly.

From the above point of view, the difference $f_0F2^+_{max} - f_0F2^-$ between the maximum crest (marked by "+") and trough ("−") of the anomaly, which was often used in the previous studies,[2] is not the optimal choice. Firstly, this parameter characterizes only the amplitude of the major crest. Secondly, if the anomaly begins to develop asymmetrically and then acquires the two-crest shape (as illustrated in the right-hand panel of Fig. 1), the above-written difference jumps (or drops) instantly by a considerable value when the second (minor) crest smoothly appears (or disappears).

A more reasonable parameter, proposed in our recent work (Dumin, 2002), is defined as

$$P = \frac{(f_0F2^+_N - f_0F2^-)(f_0F2^+_S - f_0F2^-)}{(1/2)(f_0F2^+_N + f_0F2^+_S) - f_0F2^-} \;, \qquad (1)$$

where superscripts "+" and "−" refer to the anomaly crest and trough, respectively; and subscripts "N" and "S" designate the north and south. (It is assumed here that $P = 0$ if the anomaly trough is absent.) The main advantage of $P$ is that it behaves continuously both in the cases of symmetric and asymmetric development (or decay). In addition, when structure of the anomaly is approximately symmetric, $P$ is reduced just to the difference between the amplitudes of the crest and trough.

## 3. Maps of the equatorial anomaly development

A convenient form of presentation of the equatorial anomaly development in space and time was proposed by Hopkins (1972), who drew the maps of its degree of development by the Ariel-3 satellite data as function of longitude and local time (LT).

A similar approach was applied in our work (Dumin, 2002) based on the data by ISS-b satellite; but the exactly-specified parameter $P$ was utilized instead of the qualitative estimates by Hopkins (1972), and universal time (UT) was used instead of the local time because of the specific form of the original data presentation by the ISS-b team (Atlas, 1983). The main conclusion of this work was that the tempo–longitudinal patterns of the anomaly derived from the above-mentioned data sets, in general, coincide with each other very well. This is a quite nontrivial fact, since these data were obtained in two absolutely different time periods by different instrumental methods (the top-side sounding by ISS-b and in situ probing by Ariel-3) and, thereby, refer to somewhat different altitudes.

Therefore, the above-mentioned form of description of longitudinal structure of the equatorial anomaly can serve as a convenient method for testing the ionospheric models in the low-latitude F2-layer, while the ISS-b measurements (which are one of the most numerous and systematized satellite data set available by now) can be used as a reliable primary standard.

So, to test the last version of the International Reference Ionosphere IRI-2001 (Bilitza, 2004), we drew

---

[2] For example, in the work by Deminova (1995) it was denoted by $n$.

the contours of the parameter *P* in longitude–UT coordinates, shown in Fig. 2. The algorithm of identification of the equatorial anomaly involved two basic steps:

(1) At each longitude, a minimum value of $f_0F2$ is sought for in the interval from –20° to 20° of magnetic latitude. If the minimum value of the critical frequency $f_0F2^-$ was found (e.g., at the magnetic latitude $\varphi^-$), then we should proceed to the next step. If the minimum was not found, then *P* is taken to be zero.

(2) The south crest of the anomaly is sought for in the interval from ($\varphi^- - 30°$) to $\varphi^-$; and the north crest, in the interval from $\varphi^-$ to ($\varphi^- + 30°$). If both maxima, $f_0F2^+_S$ and $f_0F2^+_N$, were uniquely determined, then the value of *P* is calculated by Eq. (1). Otherwise, if one of the maxima was not found at all, or there were multiple maxima within the above-specified intervals, then *P* is taken to be zero.

Since the critical frequency of F2-layer can be obtained in IRI-2001 by one of alternative submodels – the purely empirical CCIR or semi-empirical URSI, – both of them were used for the calculations, and the corresponding results are presented in the top and bottom left-hand panels, respectively. For the sake of comparison, the contours derived from the ISS-b satellite data are presented in the top right-hand panel.

The inclined dashed lines in Fig. 2 show the trajectories of the subsolar point (i.e. the local noon). Each panel comprises the isocontours of *P* for four periods listed in Table 1; the most of them are approximately centered to equinoxes. (The available satellite data are most abundant and uniform just around the equinoxes.) These 4-months periods are the time intervals required for collecting a complete set of worldwide data by the ISS-b satellite; and they are exactly the same as used in the original analysis by ISS-b team (for more details, see Atlas, 1983; Matuura et al., 1981; Wakai and Matuura, 1980). The corresponding isocontours for URSI and CCIR models were calculated for a single date in the center of each interval. For better visualization of longitudinal variations, the range of longitude in each map was taken to be 1.5 times greater than 360º.

## 4. Discussion

The following main conclusions can be derived from the analysis of maps presented in Fig. 2:

1. The data by ISS-b satellite demonstrate a much more fine structure in the degree of development of Appleton anomaly than both URSI and CCIR models, incorporated into IRI. (In other words, the "islands" of *P* in ISS-b maps are much more numerous, smaller in size, and greater in amplitude.) It is important to emphasize that this fine structure is in no way related to the mathematical method of interpolation of the experimental data, because the number of zonal harmonics taken into account in CCIR and URSI maps is even greater than in ISS-b maps. Besides, the same kind of the fine structure was revealed in the independent satellite data, e.g. by Ariel-3 (Hopkins, 1972) and Intercosmos-19 (Deminova, 1995; Depuev and Pulinets, 2000; Dumin and Sitnov, 1994; Karpachev, 1988; Karpachev et al., 2003).[3]

2. In general, CCIR maps reveal more fine details than URSI maps. Unfortunately, these details do not coincide with the ones in ISS-b satellite maps. So, as was already mentioned in the Introduction, the longitudinal distortions caused by the "screen" points are really substantial. Moreover, the situation is especially bad at low latitudes, because it cannot be corrected by theoretical account of only the thermospheric-wind influence on the ionospheric plasma, since the low-latitude F2-layer is strongly affected also by the **E**×**B**-drift originating in E-layer (Bilitza et al., 1987).

Therefore, as follows from the above consideration, the accuracy of description of longitudinal variations in the equatorial (Appleton) anomaly by IRI-2001 remains insufficient, and further improvements are desirable.

## Acknowledgements

We are grateful to D.K. Bilitza and B.W. Reinisch for valuable comments and suggestions to the text of this article, as well as to all developers of IRI (Bilitza, 2004) for making this model available via Internet and to the ISS-b team for publication of the digital maps of $f_0F2$ (Atlas, 1983).

---

[3] As follows from a more careful statistical analysis (Dumin, 1993), although the higher longitudinal harmonics of $f_0F2$ are really present, their statistical significance is less than for the main (first) harmonic.

**References**

Appleton, E.V. Two anomalies in the ionosphere. Nature 157, 691, 1946.
Atlas of Ionospheric Critical Frequency ($f_0$F2) Obtained from Ionosphere Sounding Satellite-b Observation, Part 4: August to December 1979 and Tables of Coefficients for Numerical Mapping of $f_0$F2. Radio Res. Labs., Ministry of Posts and Telecoms., Japan, 1983.
Bilitza, D. IRI: An international Rawer initiative. Adv. Space Res. 15(2), 7-10, 1995.
Bilitza, D. International Reference Ionosphere, http://nssdc.gsfc.nasa.gov/space/model/ionos/iri.html , 2004.
Bilitza, D., Rawer, K., Pallaschke, S., Rush, C.M., Matuura, N., Hoegy, W.R. Progress in modeling the ionospheric peak and topside electron density. Adv. Space Res. 7(6), 5-12, 1987.
Bilitza, D., Rawer, K., Bossy, L., Gulyaeva, T. International Reference Ionosphere – past, present, and future: I. Electron density. Adv. Space Res. 13(3), 3-13, 1993.
Bradley, P.A. Mapping the critical frequency of the F2-layer: Part 1 – Requirements and developments to around 1980. Adv. Space Res. 10(8), 47-56, 1990.
Chernyshev, O.V., Vasil'eva, T.N. Prognoz maksimal'nykh primenimykh chastot: W=10 (Forecast of the Maximum Usable Frequencies: W=10). Nauka, Moscow, 1973. (In Russian)
Deminova, G.F. Undulatory structure of longitudinal changes of the night equatorial anomaly. Geomagnetizm i aeronomiya 35(4), 169-173, 1995. (In Russian)
Depuev, V.H., Pulinets, S.A. Global distribution of night-time F2 peak density (Intercosmos-19 data). Adv. Space Res. 25(1), 105-108, 2000.
Dumin, Yu.V. Statistical importance of longitudinal variations of ionospheric parameters, in: Soboleva, T.N., Shashun'kina, V.M. (Eds.), Ionospheric Researches, vol. 49. Nat. Geophys. Committee, Moscow, pp. 82-89, 1993. (in Russian)
Dumin, Yu.V. Global structure of longitudinal variations in the equatorial anomaly of ionospheric F2-layer. Adv. Space Res. 29(6), 907-910, 2002.
Dumin, Yu.V., Sitnov, Yu.S. The zonal structure of the F2-region over the magnetic equator. Geomagnetizm i aeronomiya 34(4), 154-156, 1994. (In Russian)
Hopkins, H.D. Longitudinal variation of the equatorial anomaly. Planet. Space Sci. 20(12), 2093-2098, 1972.
Karpachev, A.T. The peculiarity of the global longitudinal effect of the night-time equatorial anomaly. Geomagnetizm i aeronomiya 28(1), 46-49, 1988. (In Russian)
Karpachev, A.T., Deminova, G.F., Depuev, V.H., Kochenova, N.A. Diurnal variations of the peak electron density distribution pattern at low latitudes derived from Intercosmos-19 topside sounding data. Adv. Space Res. 31(3), 521-530, 2003.
Matuura, N., Kotaki, M., Miyazaki, S., Sagawa, E., Iwamoto, I. ISS-b experimental results on global distributions of ionospheric parameters and thunderstorm activity. Acta Astronautica 8(5-6), 527-548, 1981.
Ramanamurty, Y.V. Highlights of the URSI/COSPAR workshop on IRI. Adv. Space Res. 4(1), 153-163, 1984.
Rawer, K. Modelling of neutral and ionized atmospheres, in: Flügge, S. (Ed.), Handbuch der Physik, vol. 49/7, Geophysik III, part 7. Springer, Berlin, pp. 223-535, 1984a.
Rawer, K. New description of the electron density profile. Adv. Space Res. 4(1), 11-15, 1984b.
Rawer, K. Final summary and conclusions. Adv. Space Res. 4(1), 165-169, 1984c.
Rawer, K. Problems arising in empirical modeling of the terrestrial ionosphere. Adv. Space Res. 14(12), 7-16, 1994.
Rawer, K., Bilitza, D., Ramakrishnan, S. Goals and status of the International Reference Ionosphere. Rev. Geophys. Space Phys. 16(2), 177-181, 1978.
Wakai, N., Matuura, N. Operation and experimental results of the Ionosphere Sounding Satellite-b. Acta Astronautica 7(8-9), 999-1020, 1980.

Table 1. Periods for which the ISS-b data are available; the numeration is according to the original Atlas (1983)

| No. | Dates |
|---|---|
| 1 | 11.08.1978 – 12.12.1978 |
| 2 | 10.10.1978 – 11.02.1979 |
| 4 | 10.01.1979 – 14.05.1979 |
| 6 | 08.08.1979 – 13.12.1979 |

**Figures**

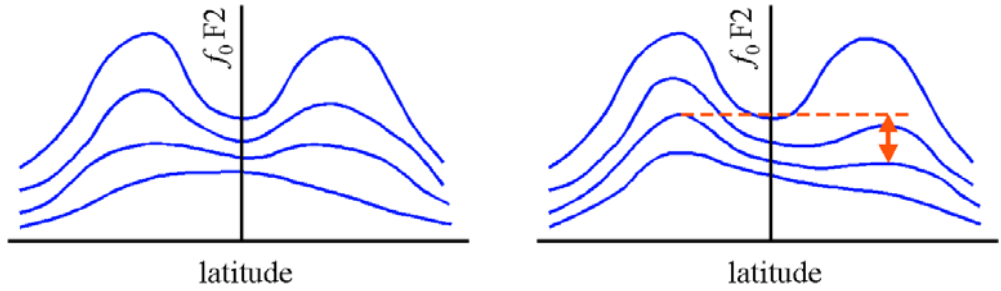

Fig. 1. A scheme of the approximately symmetric (left) and strongly asymmetric (right) development/decay of the equatorial anomaly. The thick oppositely-directed arrows in the right-hand panel indicate a sharp jump experienced by the difference $f_0F2^+_{max} - f_0F2^-$ when the minor crest of the anomaly appears/disappears.

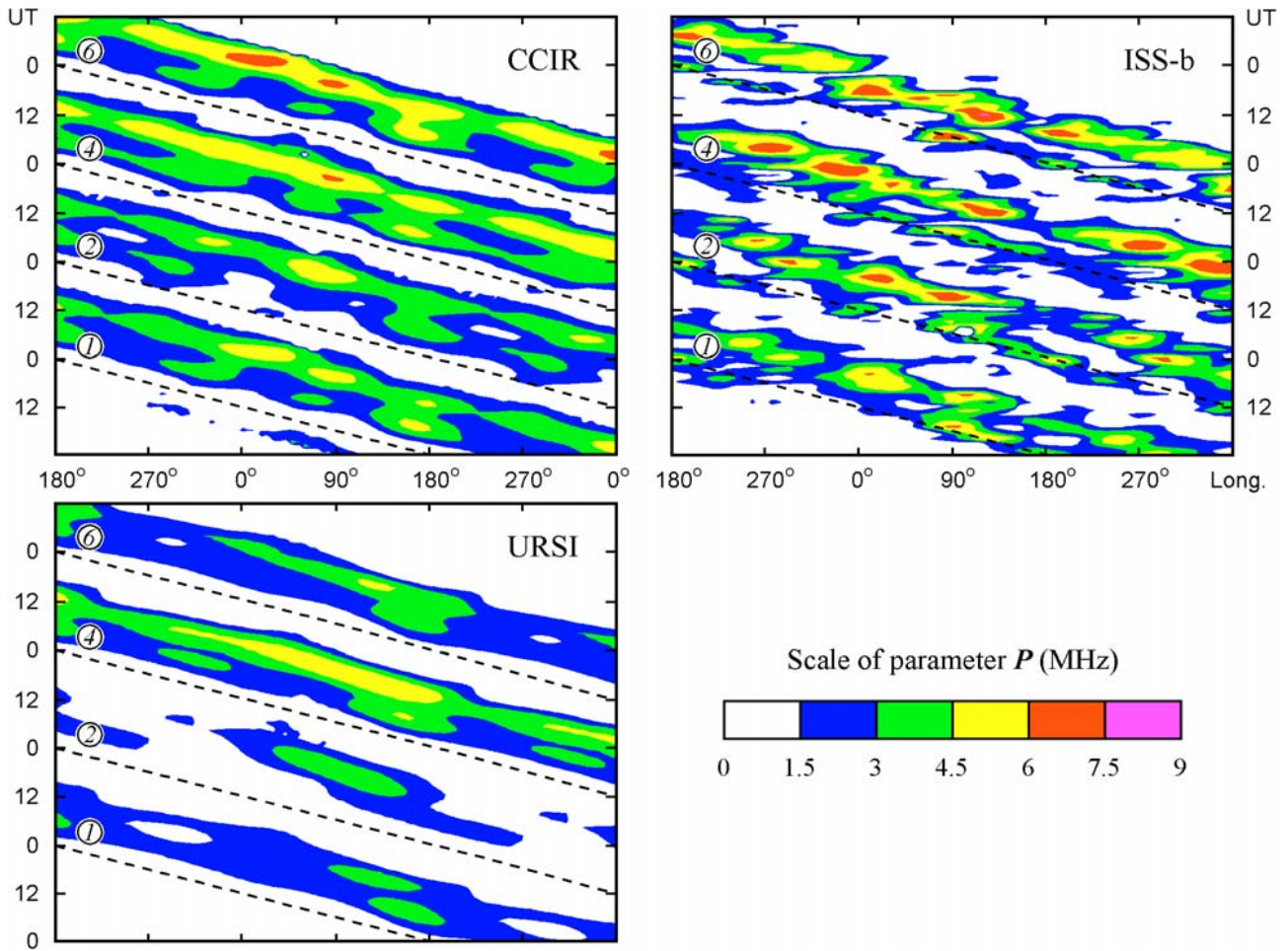

Fig. 2. Maps of the parameter $P$, characterizing the degree of development of the two-crest equatorial anomaly, derived from CCIR (top left-hand panel) and URSI (bottom left-hand panel) models as well as from ISS-b satellite data (top right-hand panel) for the time periods Nos. 1, 2, 4, and 6 (denoted by the numbers in circles, as listed in Table 1).